\documentclass[a4paper,11pt]{amsart}
\usepackage{hyperref}
\hypersetup{
    pdftitle={Lie Groups and Mechanics: an introduction},
    pdfauthor={Boris Kolev},
    pdfsubject={Mathematics}
    }
\newtheorem{theorem}{Theorem}[section]

\newtheorem{lemma}[theorem]{Lemma}

\theoremstyle{definition}

\theoremstyle{remark}
\newtheorem*{remark}{Remark}
\newtheorem*{example}{Example}
\newcommand{\Real}{\mathbb{R}}

\newcommand{\norm}[1]{\left\Vert#1\right\Vert}

\newcommand{\set}[1]{\left\{#1\right\}}

\newcommand{\commut}[2]{\left[#1,#2\,\right]}
\newcommand{\brackets}[2]{\left\langle #1,#2\,\right\rangle}
\newcommand{\pairing}[2]{\left( #1,#2\,\right)}
\newcommand{\poisson}[2]{\left\{ #1,#2\,\right\}}

\DeclareMathOperator{\tr}{tr} \DeclareMathOperator{\curl}{curl}
\DeclareMathOperator{\grad}{grad}
\DeclareMathOperator{\divergence}{div}
\begin{document}

\title[Lie Groups and Mechanics]{Lie Groups and Mechanics, \\ An introduction}%
\author{Boris Kolev}%
\address{CMI, 39, rue F. Joliot-Curie, 13453 Marseille cedex 13, France }%
\email{boris.kolev@up.univ-mrs.fr}%

%\thanks{}%
%\subjclass{}%
%\keywords{}%

%\date{}%
%\dedicatory{}%
%\commby{}%
% ----------------------------------------------------------------
\begin{abstract}
The aim of this paper is to present aspects of the use of Lie
groups in mechanics. We start with the motion of the rigid body
for which the main concepts are extracted. In a second part, we
extend the theory for an arbitrary Lie group and in a third
section we apply these methods for the diffeomorphism group of the
circle with two particular examples: the Burger equation and the
Camassa-Holm equation.
\end{abstract}
\maketitle

%--------------------------------------------------------------------

\section*{Introduction}

The aim of this article is to present aspects of the use of Lie
groups in mechanics. In a famous article \cite{Arnold66}, Arnold
showed that the motion of the rigid body and the motion of an
incompressible, inviscid fluid have the same structure. Both
correspond to the geodesic flow of a one-sided invariant metric on
a Lie group. From a rather different point of view, Jean-Marie
Souriau has pointed out in the seventies \cite{Souriau97} the
fundamental role played by Lie groups in mechanics and especially
by the dual space of the Lie algebra of the group and the
coadjoint action. We aim to discuss some aspects of these notions
through examples in finite and infinite dimension. The article is
divided in three parts. In Section~\ref{sec:motion-rigid-body} we
study in detail the motion of an $n$-dimensional rigid body. In
the second section, we treat the geodesic flow of left-invariant
metrics on an arbitrary Lie group (of finite dimension). This
permits us to extract the abstract structure from the case of the
motion of the rigid body which we presented in
Section~\ref{sec:motion-rigid-body}. Finally, in the last section,
we study the geodesic flow of $H^{k}$ right-invariant metrics on
$Diff(\mathbb{S}^{1})$, the diffeomorphism group of the circle,
using the approach developed in
Section~\ref{sec:Geodesi-flow-Group}. Two values of $k$ have
significant physical meaning in this example: $k=0$ corresponds to
the inviscid Burgers equation~\cite{Hormander97} and $k=1$
corresponds to the Camassa-Holm equation~\cite{CH93,CHH94}.

%--------------------------------------------------------------------

\section{The motion of the rigid body}\label{sec:motion-rigid-body}

\subsection{Rigid body}

In classical mechanics, a \emph{material system} $( \Sigma )$ in
the ambient space $\Real^{3}$ is described by a \emph{positive
measure} $\mu$ on $\Real^{3}$ with compact support. This measure
is called the \emph{mass distribution} of $( \Sigma )$.

\begin{itemize}
    \item If $\mu$ is proportional to the Dirac measure $\delta_{P}$,
    $( \Sigma )$ is the \emph{massive point} $P$, the multiplicative factor being the mass $m$ of the point.

    \item If $\mu$ is absolutely continuous with respect to the Lebesgue measure
    $\lambda$ on $\Real^{3}$, then the Radon-Nikodym derivative of $\mu$ with respect to $\lambda$ is the mass density of the system $( \Sigma )$.
\end{itemize}

In the \emph{Lagrangian formalism of Mechanics}, a \emph{motion}
of a material system is described by a smooth path $\varphi^{t}$
of \emph{embeddings} of the \emph{reference state} $\Sigma =
Supp(\mu)$ in the ambient space. A material system $( \Sigma )$ is
\emph{rigid} if each map $\varphi$ is the restriction to $\Sigma$
of an isometry $g$ of the Euclidean space $\Real^{3}$. Such a
condition defines what one calls a \emph{constitutive law of
motion} which restricts the space of \emph{probable} motions to
that of \emph{admissible} ones.

In the following section, we are going to study the motions of a
rigid body $(\Sigma)$ such that $\Sigma = Supp(\mu)$ spans the $3$
space. In that case, the manifold of all possible configurations
of $(\Sigma)$ is completely described by the $6$-dimensional
\emph{bundle} of frames of $\Real^{3}$, which we denote
$\mathcal{R}(\Real^{3})$. The group $D_{3}$ of
orientation-preserving isometries of $\Real^{3}$ acts simply and
transitively on that space and we can identify
$\mathcal{R}(\Real^{3})$ with $D_{3}$. Notice, however, that this
identification is not canonical -- it depends of the choice of a
"reference" frame $\Re _0$.

Although the physically meaningful rigid body mechanics is in
dimension $3$, we will not use this peculiarity in order to
distinguish easier the main underlying concepts. Hence, in what
follows, we will study the motion of an $n$-dimensional rigid
body.

Moreover, since we want to insist on concepts rather than struggle
with heavy computations, we will restrain our study to motions of
a rigid body having a fixed point. This reduction can be justified
physically by the possibility to describe the motion of an
isolated body in an \emph{inertial frame} around its \emph{center
of mass}. In these circumstances, the configuration space reduces
to the group $SO(n)$ of isometries which fix a point.

\subsection{Lie algebra of the rotation group}

The Lie algebra $\mathfrak{so}(n)$ of $SO(n)$ is the space of all
skew-symmetric $n \times n$ matrices\footnote{ In dimension $3$,
we generally identify the Lie algebra $\mathfrak{so}(3)$ with
$\Real^{3}$ endowed with the Lie bracket given by the cross
product $\omega_{1} \times \omega_{2}$.}. There is a canonical
inner product, the so-called \emph{Killing form} \cite{Souriau97}
\begin{equation*}
    \brackets{\Omega_{1}}{\Omega_{2}} = - \frac{1}{2} \,
\tr(\Omega_{1}\Omega_{2})
\end{equation*}
which permit us to identify $\mathfrak{so}(n)$ with its dual space
$\mathfrak{so}(n)^{*}$.

For $x$ and $y$ in $\Real^{n}$, we define
\begin{equation*}
    L^{*}(x,y)(\Omega) = (\Omega \, x) \cdot y ,\quad \Omega \in \mathfrak{so}(n)
\end{equation*}
which is skew-symmetric in $x,y$ and defines thus a linear map
\begin{equation*}
    L^{*}:\bigwedge^{2}\Real^{n} \rightarrow \mathfrak{so}(n)^{*} \; .
\end{equation*}
This map is injective and is therefore an isomorphism between
$\mathfrak{so}(n)^{*}$ and $\bigwedge^{2}\Real^{n}$, which have
the same dimension. Using the identification of
$\mathfrak{so}(n)^{*}$ with $\mathfrak{so}(n)$, we check that the
element $L(x,y)$ of $\mathfrak{so}(n)$ corresponding to
$L^{*}(x,y)$ is the matrix
\begin{equation}\label{equ:Loperator}
    L(x,y) = yx^{t} - xy^{t}\; .
\end{equation}
where $x^{t}$ stands for the transpose of the column vector $x$.

\subsection{Kinematics}

The location of a point $a$ of the body $\Sigma$ is described by
the column vector $r$ of its coordinates in the frame $\Re _0$. At
time $t$, this point occupies a new position $r(t)$ in space and
we have $r(t) = g(t)r$, where $g(t)$ is an element of the group
$SO(3)$. In the \emph{Lagrangian formalism}, the velocity
$\mathbf{v}(a,t)$ of point $a$ of $\Sigma$ at time $t$ is given by
\begin{equation*}
\mathbf{v}(a,t) = \frac{\partial}{\partial t}\varphi (a,t) = \dot
{g}(t)\;r.
\end{equation*}
The kinetic energy $K$ of the body $\Sigma$ at time $t$ is defined
by
\begin{equation}
K(t) = \frac{1}{2}\int_{\Sigma} \norm{\mathbf{v}(a,t)}^2 \;d\mu
     = \frac{1}{2}\int_{\Sigma} \norm{\dot{g} \, r}^{2} \;d\mu
     = \frac{1}{2}\int_{\Sigma} \norm{\Omega r}^{2} \;d\mu
\end{equation}
where $\Omega = g^{-1} \, \dot{g}$ lies in  the Lie algebra
$\mathfrak{so}(n)$.

\begin{lemma}
We have $K = - \frac{1}{2} \, \tr(\Omega J \Omega)$, where $J$ is
the symmetric matrix with entries
\begin{equation*}
    J_{ij} = \int_{\Sigma} x_{i}x_{j}
\;d\mu \; .
\end{equation*}
\end{lemma}

\begin{proof}
Let $L:\bigwedge^{2}\Real^{n} \rightarrow \mathfrak{so}(n)$ be the
operator defined by~\eqref{equ:Loperator}. We have
\begin{equation}
    L(r,\Omega \, r) =  (rr^{t})\Omega + \Omega(rr^{t})
    \qquad  \Omega \in \mathfrak{so}(n),\   r \in \Real^{n},
\end{equation}
where $rr^{t}$ is the symmetric matrix with entries $x_{i}x_{j}$.
Therefore
\begin{equation}
(\Omega \,  r) \cdot  (\Omega \,  r)
    = L^{*}(r,\Omega r)\Omega
    = - \frac{1}{2} \, \tr \big( L(r,\Omega \,  r \big) \Omega)
    = - \tr \big( \Omega(rr^{t})\Omega \big),
\end{equation}
which leads to the claimed result after integration.
\end{proof}

The kinetic energy $K$ is therefore a positive quadratic form on
the Lie algebra $\mathfrak{so}(n)$. A linear operator
$A:\mathfrak{so}(n)\rightarrow \mathfrak{so}(n)$, called the
\emph{inertia tensor} or the \emph{inertia operator}, is
associated to $K$ by means of the relation
\begin{equation*}
    K = \frac{1}{2} \brackets{A(\Omega)}{\Omega}, \qquad \Omega\in\mathfrak{so}(n).
\end{equation*}
More precisely, this operator is given by
\begin{equation}\
    A(\Omega) = J \Omega + \Omega J
              = \int_{\Sigma} \left( \Omega \, rr^{t} + rr^{t}\Omega \right)\, d\mu \; .
\end{equation}

\begin{remark}
In dimension $3$ the identification between a skew-symmetric
matrix $\Omega$ and a vector $\omega$ is given by $\omega_{1}=
-\Omega_{23}$, $\omega_{2}= \Omega_{13}$ and $\omega_{3}=
-\Omega_{12}$. If we look for a symmetric matrix $I$ such that
$A(\Omega)$ correspond to the vector $I\omega$, we find that
\begin{equation*}
I = \int_{\Sigma}\left(%
\begin{array}{ccc}
  y^{2}+z^{2} & -xy & -xz \\
  -xy & x^{2}+z^{2} & -yz \\
  -xz & -yz & x^{2}+y^{2} \\
\end{array}%
\right)d\mu ,
\end{equation*}
which gives the formula used in Classical Mechanics. $\lozenge$
\end{remark}

\subsection{Angular momentum}

In classical mechanics, we define the \emph{angular momentum} of
the body as the following $2$-vector\footnote{ In the Euclidean
$3$-space, $2$-vectors and $1$-vectors coincide. This is why,
usually, one consider the angular momentum as a $1$-vector.}
\begin{equation*}
    \mathcal{M}(t) = \int_{\Sigma} (gr) \wedge (\dot{g}r) \;d\mu\; .
\end{equation*}

\begin{lemma}
We have $L(\mathcal{M}) = gA(\Omega)g^{-1}$.
\end{lemma}

\begin{proof}
A straightforward computation shows that
\begin{equation*}
    L(gr,\dot{g}r) =  g\Omega rr^{t}g^{-1} + grr^{t}\Omega g^{-1}.
\end{equation*}
Hence
\begin{equation*}
L(\mathcal{M}) = \int_{\Sigma} L(gr,\dot{g}r) \; d\mu =
gA(\Omega)g^{-1} \;.
\end{equation*}
\end{proof}

\subsection{Equation of motion}

If there are no external actions on the body, the spatial angular
momentum is a constant of the motion,
\begin{equation}
\frac{d\mathcal{M}}{dt} = 0 \; .
\end{equation}
Coupled with the relation $L(\mathcal{M}) = gA(\Omega)g^{-1}$, we
deduce that
\begin{equation}\label{equ:Euler}
A(\dot{\Omega}) = A(\Omega) \Omega - \Omega A(\Omega)
\end{equation}
which is the generalization in $n$ dimensions of the traditional
\emph{Euler equation}. Notice that if we let $M = A(\Omega)$, this
equation can be rewritten as
\begin{equation}\label{equ:EulerDualForm}
\dot{M} = \commut{M}{\Omega}.
\end{equation}

\subsection{Integrability}

Equation \eqref{equ:EulerDualForm} has the peculiarity that the
eigenvalues of the matrix $M$ are preserved in time. Usually,
integrals of motion help to integrate a differential equation. The
\emph{Lax pairs} technique \cite{Lax68} is a method to generate
such integrals. Let us summarize briefly this technique for finite
dimensional vector spaces. Let $\dot{u} = F(u)$ be an ordinary
differential equation in a vector space $E$. Suppose that we were
able to find a smooth map $L:E\rightarrow End(F)$, where $F$ is
another vector space of finite dimension, with the following
property: if $u(t)$ is a solution of $\dot{u} = F(u)$, then the
operators $L(t) = L(u(t))$ remain conjugate with each other, that
is, there is a one-parameter family of invertible operators $P(t)$
such that
\begin{equation}\label{equ:LaxConjugate}
    L(t) = P(t)^{-1}L(0)P(t)  \; .
\end{equation}
In that case, differentiating \eqref{equ:LaxConjugate}, we get
\begin{equation}\label{equ:Lax}
    \dot{L} = \commut{L}{B}
\end{equation}
where $B = P^{-1}\dot{P}$. Conversely, if we can find a smooth
one-parameter family of matrices $B(t) \in End(F)$, solutions of
equation~\eqref{equ:Lax}, then \eqref{equ:LaxConjugate} is
satisfied with $P(t)$ a solution of $\dot{P} = PB \; .$ If this is
the case, then the eigenvalues, the trace and more generally all
conjugacy invariants of $L(u)$ constitute a set of integrals for
$\dot{u} = F(u)$.

A \emph{Hamiltonian system} on $\Real^{2N}$ is called
\emph{completely integrable} if it has $N$ integrals \emph{in
involution} that are functionally independent almost everywhere. A
theorem of Liouville describes in that case, at least
qualitatively, the dynamics of the equation. This is the reason
why it is so important to find integrals of motions of a given
differential equation.

Using the Lax pairs technique, Manakov \cite{Manakov76} proved the
following theorem

\begin{theorem}
Given any $n$, equation~\eqref{equ:EulerDualForm} has
\begin{equation*}
    N(n) = \frac{1}{2} \left[  \frac{n}{2} \right] + \frac{n(n-1)}{4}
\end{equation*}
integrals of motion in involution. The equation of motion of an
$n$-dimensional rigid body is completely integrable.
\end{theorem}

\begin{proof}[Sketch of proof]
The proof is based on the following basic lemma.

\begin{lemma}
Euler's equations \eqref{equ:EulerDualForm} of the dynamics of an
$n$-dimensional rigid body have, for any $n$, a representation in
Lax's form in matrices, linearly dependent on a parameter $\lambda
\in \mathbb{C}$, given by $L_{\lambda} = M + J^{2}\lambda$ and
$B_{\lambda} = \Omega + J\lambda$.
\end{lemma}

Hence, the polynomials $P_{k}(\lambda) = tr \, ( M + J^{2}\lambda
)^{k}$, $(k=2, \dotsc ,n)$ are time-independent and the
coefficients $P_{k}(\lambda)$ are integrals of motion. Since $M$
is skew-symmetric and $J$ is symmetric, the coefficient of
$\lambda^{s}$ in $P_{k}(\lambda)$  is nonzero, provided $s$ has
the same parity as $k$. The calculation of $N(n)$ here presents no
difficulties.\end{proof}

%--------------------------------------------------------------------

\section{Geodesic flow on a Lie Group} \label{sec:Geodesi-flow-Group}

In this section, we are going to study the geodesic flow of a left
invariant metric on a Lie group of finite dimension. Our aim is to
show that all the computations performed in
Section~\ref{sec:motion-rigid-body} are a very special case of the
theory of one-sided invariant metrics on a Lie group. Later on, we
will use these techniques to handle partial differential
equations. We refer to \cite{AK98} from where materials of this
section come from and to Souriau's book \cite{Souriau97} for a
thorough discussion of the role played by the dual of the Lie
algebra in mechanics and physics.

\subsection{Lie Groups}

A Lie group $G$ is a group together with a smooth structure such
that $g \mapsto g^{-1}$ and $(g,h) \mapsto gh$ are smooth. On $G$,
we define the \emph{right translations} $R_{h}: G \rightarrow G$
by $R_{g}(h) = hg$ and the \emph{left translations} $L_{g}: G
\rightarrow G$ by $L_{g}(h) = gh$.

A Lie group is equipped with a canonical \emph{vector-valued} one
form, the so called \emph{Maurer-Cartan form} $\omega(X_{g}) =
L_{g^{-1}}X_{g}$ which shows that the tangent bundle to $G$ is
trivial $TG \simeq G \times \mathfrak{g}$. Here $\mathfrak{g}$ is
the tangent space at the group unity $e$.

A \emph{left-invariant} tensor is completely defined by its value
at the group unity $e$. In particular, there is an isomorphism
between the tangent space at the origin and left-invariant vector
fields. Since the Lie bracket of such fields is again a
left-invariant vector field, the Lie algebra structure on vector
fields is inherited by the tangent space at the origin
$\mathfrak{g}$. This space $\mathfrak{g}$ is called the \emph{Lie
algebra} of the group $G$.

\begin{remark}
One could have defined the Lie bracket on $\mathfrak{g}$ by
pulling back the Lie bracket of vector fields by right
translation. The two definitions differ just by a minus sign
\begin{equation*}
\commut{\xi}{\omega}_{R} = - \commut{\xi}{\omega}_{L}.
\qquad\lozenge
\end{equation*}
\end{remark}

\begin{example}
The Lie algebra $\mathfrak{so}(n)$ of the rotation group $SO(n)$
consists of skew-symmetric $n \times n$ matrices. $\lozenge$
\end{example}

\subsection{Adjoint representation of $G$}

The composition $I_{g} = R_{g^{-1}}L_{g}:G \rightarrow G$ which
sends any group element $h\in G$ to $ghg^{-1}$ is an automorphism,
that is,
\begin{equation*}
    I_{g}(hk) = I_{g}(h)I_{g}(k) .
\end{equation*}
It is called an \emph{inner automorphism} of $G$. Notice that
$I_{g}$ preserves the group unity.

The differential of the inner automorphism $I_{g}$ at the group
unity $e$ is called the \emph{group adjoint operator} $Ad_{g}$
defined by
\begin{equation*}
    Ad_{g}:\mathfrak{g} \rightarrow \mathfrak{g}, \qquad
    Ad_{g} \; \omega = \frac{d}{dt}|_{t=0} \; I_{g}(h(t)),
\end{equation*}
where $h(t)$ is a curve on the group $G$ such that $h(0) = e$ and
$\dot{h}(0) = \omega \in \mathfrak{g} = T_{e}G$. The \emph{orbit}
of a point $\omega$ of $\mathfrak{g}$ under the action of the
adjoint representation is called an \emph{adjoint orbit}. The
adjoint operators form a representation of the group $G$ (i.e.
$Ad_{gh} = Ad_{g}Ad_{h}$) which preserves the Lie bracket of
$\mathfrak{g}$, that is,
\begin{equation*}
\commut{Ad_{g}\; \xi}{Ad_{g}\; \omega} = Ad_{g}\;
\commut{\xi}{\omega}.
\end{equation*}
This is the \emph{Adjoint representation} of $G$ into its Lie
algebra $\mathfrak{g}$.

\begin{example}
For $g \in SO(n)$ and $\Omega \in \mathfrak{so}(n)$, we have
$Ad_{g}\; \Omega = g \Omega g^{-1}$. $\lozenge$
\end{example}

\subsection{Adjoint representation of $\mathfrak{g}$}

The map $Ad$, which associates the operator $Ad_{g}$ to a group
element $g\in G$, may be regarded as a map from the group $G$ to
the space $End(\mathfrak{g})$ of \emph{endomorphisms} of
$\mathfrak{g}$. The differential of the map $Ad$ at the group
unity is called the adjoint representation of the Lie algebra
$\mathfrak{g}$ into itself,
\begin{equation*}
    ad:\mathfrak{g} \rightarrow End(\mathfrak{g}), \qquad
    ad_{\xi} \; \omega = \frac{d}{dt}|_{t=0} \; Ad_{g(t)} \; \omega.
\end{equation*}
Here $g(t)$ is a curve on the group $G$ such that $g(0) = e$ and
$\dot{g}(0) = \xi$. Notice that the space $\set{ad_{\xi}\; \omega,
\; \xi \in \mathfrak{g}}$ is the tangent space to the adjoint
orbit of the point $\omega \in \mathfrak{g}$.

\begin{example}
On the rotation group $SO(n)$, we have $ad_{\Xi}\; \Omega =
\commut{\Xi}{\Omega}$, where $\commut{\Xi}{\Omega} = \Xi\Omega -
\Xi\Sigma$ is the commutator of the skew-symmetric matrices $\Xi$
and $\Omega$. As we already noticed, for $n=3$, the vector
$\commut{\xi}{\omega}$ is the ordinary cross product $\xi \times
\omega$ of the angular velocity vectors $\xi$ and $\omega$ in
$\mathbb{R}^{3}$. More generally, if $G$ is an arbitrary Lie group
and $\commut{\xi}{\omega}$ is the Lie bracket on $\mathfrak{g}$
defined earlier, we have $ad_{\xi}\; \omega =
\commut{\xi}{\omega}$. $\lozenge$
\end{example}

\subsection{Coadjoint representation of $G$}

Let $\mathfrak{g}^{*}$ be the dual vector space to the Lie algebra
$\mathfrak{g}$. Elements of $\mathfrak{g}^{*}$ are linear
functionals on $\mathfrak{g}$. As we shall see, the leading part
in mechanics is not played by the Lie algebra itself but by its
dual space $\mathfrak{g}^{*}$. Souriau \cite{Souriau97} pointed
out the importance of this space in physics and called the
elements of $\mathfrak{g}^{*}$ \emph{torsors} of the group $G$.
This definition is justified by the fact that torsors of the usual
group of affine Euclidean isometries of $\mathbb{R}^{3}$ represent
the \emph{torsors} or \emph{torques} of mechanicians.

Let $A: E \rightarrow F$ be a linear mapping between vector
spaces. The dual (or adjoint) operator $A^{*}$, acting in the
reverse direction between the corresponding dual spaces, $A^{*}:
F^{*} \rightarrow E^{*}$, is defined by
\begin{equation*}
    (A^{*}\; \alpha) (x) = \alpha(A\; x)
\end{equation*}
for every $x\in E$, $\alpha \in F^{*}$.

The \emph{coadjoint representation} of a Lie group $G$ in the
space $\mathfrak{g}^{*}$ is the representation that associates to
each group element $g$ the linear transformation
\begin{equation*}
    Ad^{*}_{g}:\mathfrak{g}^{*} \rightarrow \mathfrak{g}^{*}
\end{equation*}
given by $Ad^{*}_{g} = (Ad_{g^{-1}})^{*}$. In other words,
\begin{equation*}
    (Ad^{*}_{g}\; m)(\omega) = m(Ad_{g}\; \omega)
\end{equation*}
for every $g\in G$, $m \in \mathfrak{g}^{*}$ and $\omega \in
\mathfrak{g}$. The choice of $g^{-1}$ in the definition of
$Ad^{*}_{g}$ is to ensure that $Ad^{*}$ is a \emph{left
representation}, that is $Ad^{*}_{gh} = Ad^{*}_{g}Ad^{*}_{h}$ and
not the converse (or \emph{right representation}). The
\emph{orbit} of a point $m$ of $\mathfrak{g}^{*}$ under the action
of the coadjoint representation is called a \emph{coadjoint
orbit}.

The \emph{Killing form} on $\mathfrak{g}$ is defined by
\begin{equation*}
    k(\xi,\omega) = \tr\,\left( ad_{\xi} \, ad_{\omega}\right).
\end{equation*}
Notice that $k$ is invariant under the adjoint representation of
$G$. The Lie group $G$ is \emph{semi-simple} if $k$ is
non-degenerate. In that case, $k$ induces an isomorphism between
$\mathfrak{g}$ and $\mathfrak{g}^{*}$ which permutes the adjoint
and coadjoint representation. The adjoint and coadjoint
representation of a semi-simple Lie group are \emph{equivalent}.

\begin{example}
For the group $SO(3)$ the coadjoint orbits are the sphere centered
at the origin of the $3$-dimensional space $\mathfrak{so}(3)^{*}$.
They are similar to the adjoint orbits of this group, which are
spheres in the space $\mathfrak{so}(3)$. $\lozenge$
\end{example}

\begin{example}
For the group $SO(n)$ ($n\geq 3$), the adjoint representation and
coadjoint representations are equivalent due to the non-degeneracy
of the Killing form\footnote{This formula is exact up to a scaling
factor since a precise computation for $\mathfrak{so}(n)$ gives
$k(X,Y) = (n-2) \tr(XY)$.}
\begin{equation*}
    k(\Xi,\Omega) = \frac{1}{2} \, \tr\,\left( \Xi \, \Omega^{*}\right),
\end{equation*}
where $\Omega^{*}$ is the transpose of $\Omega$ relative to the
corresponding inner product of $\mathbb{R}^{n}$. Therefore
\begin{equation*}
    Ad_{g}^{*}\; M = g M g^{-1},
\end{equation*}
for $M \in \mathfrak{so}(n)^{*}$ and $g\in SO(n)$. $\lozenge$
\end{example}

Despite the previous two examples, in general the coadjoint and
the adjoint representations are not alike. For example, this is
the case for the \emph{Poincar\'{e} group} (the non-homogenous
\emph{Lorentz group}) cf. \cite{CvK03}.

\subsection{Coadjoint representation of $\mathfrak{g}$}

Similar to the adjoint representation of $\mathfrak{g}$, there is
the coadjoint representation of $\mathfrak{g}$. This later is
defined as the dual of the adjoint representation of
$\mathfrak{g}$, that is,
\begin{equation*}
    ad^{*}:\mathfrak{g} \rightarrow End(\mathfrak{g}^{*}), \qquad
    ad^{*}_{\xi} \; m = (ad_{\xi})^{*}(m) = - \frac{d}{dt}|_{t=0} \; Ad_{g(t)}^{*} \; m,
\end{equation*}
where $g(t)$ is a curve on the group $G$ such that $g(0) = e$ and
$\dot{g}(0) = \xi$.

\begin{example}
For $\Omega\in \mathfrak{so}(n)$ and $M\in \mathfrak{so}(n)^{*}$,
we have $ad^{*}_{\Omega}\, M = - \commut{\Omega}{M}$. $\lozenge$
\end{example}

Given $m \in \mathfrak{g}^{*}$, the vectors $ad^{*}_{\xi}\; m$,
with various $\xi \in \mathfrak{g}$, constitute the tangent space
to the coadjoint orbit of the point $m$.

\subsection{Left invariant metric on $G$}

A \emph{Riemannian} or \emph{pseudo-Riemannian} metric on a Lie
group $G$ is left invariant if it is preserved under every left
shift $L_{g}$, that is,
\begin{equation*}
    \brackets{X_{g}}{Y_{g}}_{g} =
    \brackets{L_{h}\,X_{g}}{L_{h}\,Y_{g}}_{hg},
    \qquad  g,h \in G .
\end{equation*}

A left-invariant metric is uniquely defined by its restriction to
the tangent space to the group at the unity, hence by a quadratic
form on $\mathfrak{g}$. To such a quadratic form on
$\mathfrak{g}$, a symmetric operator $A:\mathfrak{g} \rightarrow
\mathfrak{g}^{*}$ defined by
\begin{equation*}
    \brackets{\xi}{\omega} = \pairing{A\xi}{\omega} =
    \pairing{A\omega}{\xi}, \qquad  \xi ,\omega \in
    \mathfrak{g} \, ,
\end{equation*}
is naturally associated, and conversely\footnote{The round
brackets correspond to the natural pairing between elements of
$\mathfrak{g}$ and $\mathfrak{g}^{*}$.}. The operator $A$ is
called the \emph{inertia operator}. $A$ can be extended to a
left-invariant tensor $A_{g}:T_{g}G \rightarrow T_{g}G^{*}$
defined by $A_{g} = L_{g^{-1}}^{*} A L_{g^{-1}}$. More precisely,
we have
\begin{equation*}
    \brackets{X}{Y}_{g} = \pairing{A_{g}X}{Y}_{g} =
    \pairing{A_{g}Y}{X}_{g}, \qquad  X,Y \in
    T_{g}G .
\end{equation*}

The \emph{Levi-Civita} connection of a left-invariant metric is
itself left-invariant: if $L_{a}$ and $L_{b}$ are left-invariant
vector fields, so is $\nabla_{L_{a}}L_{b}$. We can write down an
expression for this connection using the operator $B: \mathfrak{g}
\times \mathfrak{g} \rightarrow \mathfrak{g}$ defined by
\begin{equation}
    \brackets{ \commut{a}{b} }{c} =\brackets{B(c,a)}{b}
\end{equation}
for every $a,b,c$ in $\mathfrak{g}$. An exact expression for $B$
is
\begin{equation*}
    B(a,b) = A^{-1}\, ad^{*}_{b}(A\, a) \, .
\end{equation*}
With these definitions, we get
\begin{equation}
    (\nabla_{L_{a}}L_{b})(e) = \frac{1}{2}\commut{a}{b} -
    \frac{1}{2} \{ B(a,b) + B(b,a) \}
\end{equation}

\subsection{Geodesics}

Geodesics are defined as extremals of the \emph{Lagrangian}
\begin{equation}
    \mathcal{L}(g) = \int K\left( g(t),\dot{g}(t) \right)\,dt
\end{equation}
where
\begin{equation}\label{equ:KineticEnergy}
    K(X) = \frac{1}{2}\, \brackets{X_{g}}{X_{g}}_{g}
         = \frac{1}{2}\, \pairing{A_{g}\,X_{g}}{X_{g}}_{g}
\end{equation}
is called the \emph{kinetic energy} or \emph{energy functional}.

If $g(t)$ is a geodesic, the velocity $\dot{g}(t)$ can be
translated to the identity via left or right shifts and we obtain
two elements of the Lie algebra $\mathfrak{g}$,
\begin{equation*}
    \omega_{L} = L_{g^{-1}}\dot{g} , \qquad \omega_{R} =
    R_{g^{-1}}\dot{g},
\end{equation*}
 called the \emph{left angular velocity}, respectively the
\emph{right angular velocity}. Letting $m = A_{g}\,\dot{g} \in
T_{g}G^{*}$, we define the \emph{left angular momentum} $m_{L}$
and the \emph{right angular momentum} $m_{R}$ by
\begin{equation*}
      m_{L} = L_{g}^{*}m \in \mathfrak{g}^{*}  , \qquad m_{R} =
    R_{g}^{*}m \in \mathfrak{g}^{*}.
\end{equation*}
Between these four elements, we have the relations
\begin{equation}
 \omega_{R} = Ad_{g}\; \omega_{L}, \quad m_{R} = Ad_{g}^{*}m_{L},
 \quad  m_{L} = A\, \omega_{L}.
\end{equation}
Note that the \emph{kinetic energy} is given by the formula
\begin{equation}
    K = \frac{1}{2}\brackets{\dot{g}}{\dot{g}}_{g}
      = \frac{1}{2}\brackets{\omega_{L}}{\omega_{L}}
      = \frac{1}{2}\pairing{m_{L}}{\omega_{L}}
      = \frac{1}{2}\pairing{A_{g}\;\dot{g}}{\dot{g}}_{g} \, .
\end{equation}

\begin{example}
The kinetic energy of an $n$-dimensional rigid body, defined by
\begin{equation}\label{equ:EulerLagrange}
K(t) = \frac{1}{2}\int_{\Sigma} \norm{\dot{g} \, r}^{2} \;d\mu
     =  - \frac{1}{2} \, \tr(\Omega J \Omega)
\end{equation}
is clearly a left-invariant Riemannian metric on $SO(n)$. In this
example, we have $\Omega = \omega_{L}$ and $M = m_{L}$.
Physically, the left-invariance is justified by the fact that the
physics of the problem must not depend on a particular choice of
reference frame used to describe it. It is a special case of
Galilean invariance. $\lozenge$
\end{example}

\subsection{Euler-Arnold equation}

The invariance of the energy with respect to left translations
leads to the existence of a \emph{momentum map} $\mu:
TG\rightarrow \mathfrak{g}^{*}$ defined by
\begin{equation*}
    \mu((g,\dot{g}))(\xi)
    = \frac{\partial K}{\partial \dot{g}} \, Z_{\xi}
    = \brackets{\dot{g}}{R_{g}\,\xi}_{g}
    = \pairing{m}{R_{g}\,\xi}
    = \pairing{ R^{*}_{g}\, m}{\xi}
    = m_{R}(\xi),
\end{equation*}
where $Z_{\xi}$ is the right-invariant vector field generated by
$\xi \in \mathfrak{g}$. According to Noether's theorem
\cite{Souriau97}, this map is constant along a geodesic, that is
\begin{equation}\label{equ:Noether}
    \frac{dm_{R}}{dt} = 0.
\end{equation}
As we did in the special case of the group $SO(n)$, using the
relation $m_{R} = Ad_{g}^{*}\, m_{L}$ and computing the time
derivative, we obtain
\begin{equation}\label{equ:EulerArnoldDualForm}
    \frac{dm_{L}}{dt} = ad^{*}_{\omega_{L}}\,m_{L} .
\end{equation}
This equation is known as the \emph{Arnold-Euler equation}. Using
$\omega_{L} = A^{-1}\, m_{L}$, it can be rewritten as an evolution
equation on the Lie algebra
\begin{equation}\label{equ:EulerArnold}
    \frac{d\omega_{L}}{dt} = B(\omega_{L},\omega_{L})\, .
\end{equation}

\begin{remark}
The \emph{Euler-Lagrange} equations of
problem~\eqref{equ:EulerLagrange} are given by
\begin{equation}\label{equ:EulerLagrangeBis}
    \left\{%
\begin{array}{ll}
    \dot{g} & = L_{g} \, \omega_{L} \, ,\\
    \dot{\omega}_{L} & = B(\omega_{L},\omega_{L}) \, .  \\
\end{array}%
\right.
\end{equation}
If the metric is bi-invariant, then $B(a,b) = 0$ for all $a,b \in
\mathfrak{g}$ and $\omega_{L}$ is constant. In that special case,
geodesics are one-parameter subgroups, as expected. $\lozenge$
\end{remark}

\subsection{Lie-Poisson structure on $\mathfrak{g}^{*}$}

A \emph{Poisson structure} on a manifold $M$ is a skew-symmetric
bilinear function $\poisson{}{}$ that associates to a pair of
smooth functions on the manifold a third function, and which
satisfies the \emph{Jacobi identity}
\begin{equation*}
    \poisson{\poisson{f}{g}}{h} + \poisson{\poisson{g}{h}}{f} +
    \poisson{\poisson{h}{f}}{g} = 0
\end{equation*}
as well as the \emph{Leibniz identity}
\begin{equation*}
    \poisson{f}{gh} = \poisson{f}{g}h + g\poisson{f}{h} .
\end{equation*}

On the torsor space $\mathfrak{g}^{*}$ of a Lie group $G$, there
is a natural Poisson structure defined by
\begin{equation}\label{equ:LiePoisson}
    \poisson{f}{g}(m) = (m,\commut{d_{m}f}{d_{m}g})
\end{equation}
for $m\in\mathfrak{g}^{*}$ and $f,g \in
C^{\infty}(\mathfrak{g}^{*})$. Note that the differential of $f$
at each point $m\in \mathfrak{g}^{*}$ is an element of the Lie
algebra $\mathfrak{g}$ itself. Hence, the commutator
$\commut{d_{m}\, f}{d_{m}\, g}$ is also a vector of this Lie
algebra. The operation defined above is called the \emph{natural
Lie-Poisson structure} on the dual space to a Lie algebra. For
more materials on Poisson structures, we refer to
\cite{MR99,Weinstein83}.

\begin{remark}
A Poisson structure on a vector space $E$ is \emph{linear} if the
Poisson bracket of two linear functions is itself a linear
function. This property is satisfied by the Lie-Poisson bracket on
the torsors space $\mathfrak{g}^{*}$ of a Lie group $G$.
$\lozenge$
\end{remark}

To each function $H$ on a Poisson manifold $M$ one can associate a
vector field $\xi_{H}$ defined by
\begin{equation*}
    L_{\xi_{H}}\, f = \poisson{H}{f}
\end{equation*}
and called the \emph{Hamiltonian field} of $H$. Notice that
\begin{equation*}
 \commut{\xi_{F}}{\xi_{H}} = \xi_{\poisson{F}{H}}.
\end{equation*}
Conversely, a vector field $v$ on a Poisson manifold is said to be
\emph{Hamiltonian} if there exists a function $H$ such that $v =
\xi_{H}$.

\begin{example}
On the torsors space $\mathfrak{g}^{*}$ of a Lie group $G$, the
Hamiltonian field of a function $H$ for the natural Lie-Poisson
structure is given by $\xi_{H}(m) = ad^{*}_{d_{m}H}\, m \, .$

Let $A$ be the inertia operator associated to a left-invariant
metric on $G$. Then equation~\eqref{equ:EulerArnoldDualForm} on
$\mathfrak{g}^{*}$ is Hamiltonian with quadratic Hamiltonian
\begin{equation*}
    H(m) = \frac{1}{2}\, \pairing{A^{-1}\, m}{M}, \qquad m\in\mathfrak{g}^{*},
\end{equation*}
which is nothing else but the kinetic energy expressed in terms of
$m=A\, \omega$. Notice that since $m_{L}(t) = Ad_{g(t)}^{*}\,
m_{R}$ where $m_{R}\in \mathfrak{g}^{*}$ is a constant, each
integral curve $m_{L}(t)$ of this equation stays on a coadjoint
orbit. $\lozenge$
\end{example}

A Poisson structure on a manifold $M$ is non-degenerate if it
derives from a \emph{symplectic structure} on $M$. That is
\begin{equation*}
    \poisson{f}{g} = \omega(\xi_{f},\xi_{g}) \, ,
\end{equation*}
where $\omega$ is a non-degenerate closed two form on $M$.
Unfortunately, the Lie-Poisson structure on $\mathfrak{g}^{*}$ is
degenerate in general. However, the restriction of this structure
on each coadjoint orbit is non-degenerate. The symplectic
structure on each coadjoint orbit is known as the
Kirillov\footnote{Jean-Marie Souriau has generalized this
construction for other natural $G$-actions on $\mathfrak{g}^{*}$
when the group $G$ has non null symplectic cohomology
\cite{Souriau97}.} form. It is given by
\begin{equation*}
    \omega(ad^{*}_{a}\, m,ad^{*}_{b}\, m) = \pairing{m}{\commut{a}{b}}
\end{equation*}
where $a,b\in \mathfrak{g}$ and $m\in\mathfrak{g}^{*}$. Recall
that the tangent space to the coadjoint orbit of $m\in
\mathfrak{g}^{*}$ is spanned by the vectors $ad^{*}_{\xi}\, m$
where $\xi$ describes $\mathfrak{g}$.

%--------------------------------------------------------------------

\section{Right-invariant metric on the diffeomorphism group}
\label{sec:Right-Invaria-metric}

In \cite{Arnold66}, Arnold showed that Euler equations of an
incompressible fluid may be viewed as the geodesic flow of a
right-invariant metric on the group of volume-preserving
diffeomorphism of a $3$-dimensional Riemannian manifold $M$
(filled by the fluid). More precisely, let $G = Diff_{\mu}(M)$ be
the group of diffeomorphisms preserving a volume form $\mu$ on
some closed Riemannian manifold $M$. According to the Action
Principle, motions of an ideal (incompressible and inviscid) fluid
in $M$ are geodesics of a right-invariant metric on
$Diff_{\mu}(M)$. Such a metric is defined by a quadratic form $K$
(the kinetic energy) on the Lie algebra $\mathcal{X}_{\mu}(M)$ of
divergence-free vector fields
\begin{equation*}
    K = \frac{1}{2}\, \int_{M}\norm{v}^{2}d\mu \,
\end{equation*}
where $\norm{v}^{2}$ is the square of the Riemannian length of a
vector field $v \in \mathcal{X}(M)$. An operator $B$ on
$\mathcal{X}_{\mu}(M) \times \mathcal{X}_{\mu}(M) $ defined by the
relation
\begin{equation*}
    \brackets{ \commut{u}{v} }{w} =\brackets{B(w,u)}{v}
\end{equation*}
exists. It is given by the formula
\begin{equation*}
    B(u,v) = \curl u \times v + \grad p \, ,
\end{equation*}
where $\times$ is the cross product and $p$ a function on $M$
defined uniquely (modulo an additive constant) by the condition
$\divergence B =0$ and the tangency of $B(u,v)$ to $\partial M$.
The Euler equation for ideal hydrodynamics is the evolution
equation
\begin{equation}
    \frac{\partial u}{\partial t} = u \times \curl u - \grad p
    \, .
\end{equation}
If at least formally, the theory works as well in infinite
dimension and the unifying concepts it brings form a beautiful
piece of mathematics, the details of the theory are far from being
as clear as in finite dimension. The main reason of these
difficulties is the fact that the diffeomorphism group is just a
\emph{Fr\'{e}chet Lie group}, where the main theorems of differential
geometry like the Cauchy-Lipschitz theorem and the Inverse
function theorem are no longer valid.

In this section, we are going to apply the results of
Section~\ref{sec:Geodesi-flow-Group} to study the geodesic flow of
a $H^{k}$ right-invariant metrics on the diffeomorphism group of
the circle $\mathbb{S}^{1}$. This may appear to be less ambitious
than to study the $3$-dimensional diffeomorphism group. However,
we will be able to understand in that example some phenomena which
may lead to understand why the $3$-dimensional ideal hydrodynamics
is so difficult to handle. Moreover, we shall give an example
where things happen to work well, the Camassa-Holm equation.

\subsection{The diffeomorphism group of the circle}

The group $Diff(\mathbb{S}^{1})$ is an open subset of
$C^{\infty}(\mathbb{S}^{1},\mathbb{S}^{1})$ which is itself a
closed subset of $C^{\infty}(\mathbb{S}^{1},\mathbb{C})$. We
define a local chart $(U_{0}, \Psi_{0})$ around a point
$\varphi_{0}\in Diff(\mathbb{S}^{1})$ by the neighborhood
\begin{equation*}
U_{0}=\set{\Vert \varphi-\varphi_0
\Vert_{C^0(\mathbb{S}^{1})}<1/2}
\end{equation*}
of $\varphi_0$ and the map
\begin{equation*}
\Psi_{0}(\varphi) = \frac{1}{2\pi i} \log ( \overline{\varphi_{0}
(x)} \varphi (x)) = u(x),\quad x \in \mathbb{S}^{1}.
\end{equation*}
The structure described above endows $Diff(\mathbb{S}^{1})$ with a
smooth manifold structure based on the Fr\'{e}chet space
$C^\infty(\mathbb{S}^{1})$. The composition and the inverse are
both smooth maps $Diff(\mathbb{S}^{1})\times
Diff(\mathbb{S}^{1})\rightarrow Diff(\mathbb{S}^{1})$,
respectively $Diff(\mathbb{S}^{1})\rightarrow
Diff(\mathbb{S}^{1})$, so that $Diff(\mathbb{S}^{1})$ is a Lie
group.

A tangent vector $V$ at a point $\varphi \in Diff(\mathbb{S}^{1})$
is a function $V:\mathbb{S}^{1}\rightarrow T\mathbb{S}^{1}$ such
that $\pi(V(x)) = \varphi(x)$. It is represented by a pair
$(\varphi , v) \in Diff(\mathbb{S}^{1})\times
C^{\infty}(\mathbb{S}^{1})$. Left and right translations are
smooth maps and their derivatives at a point $\varphi \in
Diff(\mathbb{S}^{1})$ are given by
\begin{gather*}
    L_{\psi}\, V  = (\psi(\varphi),\psi_{x}(\varphi) \, v) \\
    R_{\psi}\, V  = (\varphi(\psi),v(\psi))
\end{gather*}
The adjoint action on $\mathfrak{g} = Vect(\mathbb{S}^{1}) \equiv
C^{\infty}(\mathbb{S}^{1})$ is
\begin{equation*}
    Ad_{\psi}\, u  = \psi_{x}(\psi^{-1})u(\psi^{-1}),
\end{equation*}
whereas the Lie bracket on the Lie algebra
$T_{Id}\,{Diff(\mathbb{S}^{1})} = Vect(\mathbb{S}^{1}) \equiv
C^\infty(\mathbb{S}^{1})$ of $Diff(\mathbb{S}^{1})$ is given by
\begin{equation*}
[u,v]=-(u_xv-uv_x),\qquad u,v \in C^\infty(\mathbb{S}^{1})
\end{equation*}

Each $v \in Vect(\mathbb{S}^{1})$ gives rise to a one-parameter
subgroup of diffeomorphisms $\{ \eta(t,\cdot)\}$ obtained by
solving
\begin{equation}\label{equ:OneParameterSubGroup}
\eta_t=v(\eta)\quad\hbox{in}\quad C^\infty(\mathbb{S}^{1})
\end{equation}
with initial data $\eta(0)=Id \in Diff(\mathbb{S}^{1})$.
Conversely, each one-parameter subgroup $t \mapsto \eta(t) \in
Diff(\mathbb{S}^{1})$ is determined by its infinitesimal generator
\begin{equation*}
v=\frac{\partial}{\partial t}\, \eta(t) \Bigl|_{t=0} \in
Vect(\mathbb{S}^{1}).
\end{equation*}
Evaluating the flow $t \mapsto \eta(t,\cdot)$ of
\eqref{equ:OneParameterSubGroup} at $t=1$ we obtain an element
$\exp_L(v)$ of $Diff(\mathbb{S}^{1})$. The Lie-group exponential
map $v \to \exp_L(v)$ is a smooth map of the Lie algebra to the
Lie group \cite{Milnor84}. Although the derivative of $\exp_L$ at
$0 \in C^\infty(\mathbb{S}^{1})$ is the identity, $\exp_L$ is not
locally surjective \cite{Milnor84}. This failure, in contrast with
the case of Hilbert Lie groups \cite{Lang99}, is due to the fact
that the inverse function theorem does not necessarily hold in
Fr\'{e}chet spaces \cite{Hamilton82}.

\subsection{$H^{k}$ metrics on $Diff(\mathbb{S}^{1})$}

For $k \geq 0$ and  $u,v \in Vect(\mathbb{S}^{1}) \equiv
C^{\infty}(\mathbb{S}^{1})$, we define
\begin{equation}\label{equ:MetricHk}
\langle u,\, v \rangle_k =\int_{\mathbb{S}^{1}}
\sum_{i=0}^{k}(\partial_x^i u)\,(\partial_x^i v)\,dx =
\int_{\mathbb{S}^{1}} A_k(u)\, v\, dx \, ,
\end{equation}
where
\begin{equation}\label{equ:Ak}
A_k = 1-\frac{d^2}{dx^2}+...+(-1)^k \frac{d^{2k}}{dx^{2k}}
\end{equation}
is a continuous linear isomorphism of
$C^{\infty}(\mathbb{S}^{1})$. Note that $A_{k}$ is a symmetric
operator for the $L^{2}$ inner product
\begin{equation*}
\int_{\mathbb{S}^{1}} A_k(u)\, v\, dx = \int_{\mathbb{S}^{1}} u\,
A_k(v)\, dx.
\end{equation*}

\begin{remark}
What should be $\mathfrak{g}^{*}$ for $G=Diff(\mathbb{S}^{1})$ and
$\mathfrak{g} = vect(\mathbb{S}^{1})$ ? If we let
$\mathfrak{g}^{*}$ be the \emph{space of distributions}, $A_{k}$
is no longer an isomorphism. This is the reason why we restrict
$\mathfrak{g}^{*}$ to the range of $A_{k}$
\begin{equation*}
    Im(A_{k}) = C^{\infty}(\mathbb{S}^{1}).
\end{equation*}
The pairing between $\mathfrak{g}$ and $\mathfrak{g}^{*}$ is then
given by the $L^{2}$ inner product
\begin{equation*}
    \pairing{m}{u} = \int_{\mathbb{S}^{1}} mu\, dx .
\end{equation*}
With these definitions, the coadjoint action of
$Diff(\mathbb{S}^{1})$ on $\mathfrak{g}^{*} =
C^{\infty}(\mathbb{S}^{1})$ is given by
\begin{equation*}
    Ad^{*}_{\varphi}\, m =
    \frac{1}{(\varphi_{x}(\varphi^{-1}))^{2}}m(\varphi^{-1}).
\end{equation*}
Notice that this formula corresponds exactly to the action of the
diffeomorphism group $Diff(\mathbb{S}^{1})$ on \emph{quadratic
differentials} of the circle (expressions of the form $m(x)\,
dx^{2}$). This is the reason why one generally speaks of the
torsor space of the group $Diff(\mathbb{S}^{1})$ as the space of
quadratic differentials. $\lozenge$
\end{remark}

We obtain a smooth right-invariant metric on
$Diff(\mathbb{S}^{1})$ by extending the inner product
\eqref{equ:MetricHk} to each tangent space
$T_{\varphi}\,Diff(\mathbb{S}^{1})$, $\varphi\in
Diff(\mathbb{S}^{1})$, by right-translations i.e.
\begin{equation*}
\brackets{V}{W}_{\varphi} = \brackets{ R_{\varphi^{-1}} V }{
R_{\varphi^{-1}} W }_k,\quad V,W \in
T_{\varphi}{Diff(\mathbb{S}^{1})}.
\end{equation*}

The existence of a connection compatible with the metric is
ensured (see \cite{CK02}) by the existence of a bilinear operator
$B: C^{\infty}(\mathbb{S}^{1}) \times C^{\infty}(\mathbb{S}^{1})
\to C^{\infty}(\mathbb{S}^{1})$ such that
\begin{equation*}
\langle B(u,v),\, w \rangle=\langle u,\, [v,w] \rangle,\qquad
u,v,w \in Vect(\mathbb{S}^{1})=C^{\infty}(\mathbb{S}^{1}).
\end{equation*}
For the $H^k$ metric, this operator is given by (see \cite{CK03})
\begin{equation}
B_k(u,v)=-\,A_k^{-1}\Bigl( 2v_x A_k(u)+v A_k(u_x)\Bigr),\qquad u,v
\in  C^{\infty}(\mathbb{S}^{1}).
\end{equation}

\subsection{Geodesics}

The existence of the connection $\nabla^k$ enables us to define
the geodesic flow. A $C^2$-curve $\varphi:I \to
Diff(\mathbb{S}^{1})$ such that
$\nabla_{\dot{\varphi}}\,\dot{\varphi}=0$, where $\dot{\varphi}$
denotes the time derivative $\varphi_{t}$ of $\varphi$, is called
a \textit{geodesic}. As we did in
Section~\ref{sec:Geodesi-flow-Group}, in the case of a
left-invariant metric, we let
\begin{equation*}
u(t) = R_{\varphi^{-1}}\, \dot{\varphi}
     = \varphi_t \circ \varphi^{-1}
\end{equation*}
which is the right angular velocity on the group
$Diff(\mathbb{S}^{1})$. Therefore, a curve $\varphi \in
C^2(I,Diff(\mathbb{S}^{1}))$ with $\varphi(0)=Id$ is a geodesic if
and only if
\begin{equation}\label{equ:EulerHk}
u_t=B_k(u,u), \qquad  t \in I.
\end{equation}
Equation \eqref{equ:EulerHk} is the \textit{Euler-Arnold equation}
associated to the right-invariant metric~\eqref{equ:MetricHk}.
Here are two examples of problems of type \eqref{equ:EulerHk} on
$Diff(\mathbb{S}^{1})$ which arise in mechanics.

\begin{example}
For $k=0$, that is for the $L^{2}$ right-invariant metric,
equation \eqref{equ:EulerHk} becomes the inviscid Burgers equation
\begin{equation}\label{equ:Burgers}
    u_t+3uu_x=0.
\end{equation}
All solutions of \eqref{equ:Burgers} but the constant functions
have a finite life span and \eqref{equ:Burgers} is a simplified
model for the occurrence of shock waves in gas dynamics (see
\cite{Hormander97}). $\lozenge$
\end{example}

\begin{example}
For $k=1$, that is for the $H^{1}$ right-invariant metric,
equation \eqref{equ:EulerHk} becomes the Camassa-Holm equation
(cf. \cite{Misiolek98})
\begin{equation}\label{equ:CamassaHolm}
    u_{t} +u u_{x}+ \partial_{x} \, (1-\partial_{x^{2}})^{-1}\left( u^{2}+
\frac{1}{2}\, u_{x^{2}}\right)=0.
\end{equation}
Equation \eqref{equ:CamassaHolm} is a model for the unidirectional
propagation of shallow water waves \cite{CH93,Johson02}. It has a
bi-Hamiltonian structure \cite{FF81} and is completely integrable
\cite{CMcK99}. Some solutions of \eqref{equ:CamassaHolm} exist
globally in time \cite{Constantin97,CE98}, whereas others develop
singularities in finite time \cite{CE98,CE98b,CE00,McKean98}. The
blowup phenomenon can be interpreted as a simplified model for
wave breaking -- the solution (representing the water's surface)
stays bounded while its slope becomes unbounded \cite{CE00}.
$\lozenge$
\end{example}

\subsection{The momentum}

As a consequence of the right-invariance of the metric by the
action of the group on itself, we obtain the conservation of the
\emph{left angular momentum} $m_{L}$ along a geodesic $\varphi$.
Since $m_{L} = Ad^{*}_{\varphi^{-1}}\, m_{R}$ and $m_{R} =
A_{k}(u)$, we get that
\begin{equation}\label{equ:Momentum}
    m_k(\varphi , t)=A_{k}(u) \circ \varphi \cdot \varphi_{x}^{2},
\end{equation}
satisfies $m_k(t)=m_k(0)$ as long as $m_k(t)$ is defined.

\subsection{Existence of the geodesics}

In a local chart the geodesic equation \eqref{equ:EulerHk} can be
expressed as the Cauchy problem
\begin{equation}\label{equ:Cauchy}
 \left \{
  \begin{array}{ccc}
    \varphi_{t} & = & v,\\
    v_{t} & = & P_{k}(\varphi,v),
  \end{array}
 \right .
\end{equation}
with $\varphi(0)=Id,\,v(0)=u(0)$. However, the local existence
theorem for differential equations with smooth right-hand side,
valid for Hilbert spaces \cite{Lang99}, does not hold in $C^\infty
(\mathbb{S}^{1})$ (see \cite{Hamilton82}) and we cannot conclude
at this stage. However, in~\cite{CK03}, we proved

\begin{theorem}\label{thm:Theorem2}
Let $k \ge 1$. For every $u_0 \in C^\infty (\mathbb{S}^{1})$,
there exists a unique geodesic $\varphi \in
C^\infty([0,T),Diff(\mathbb{S}^{1}))$ for the metric
\eqref{equ:MetricHk}, starting at $\varphi(0)=Id \in
Diff(\mathbb{S}^{1})$ in the direction $u_0=\varphi_t(0) \in
Vect(\mathbb{S}^{1})$. Moreover, the solution depends smoothly on
the initial data $u_0\in C^\infty (\mathbb{S}^{1})$.
\end{theorem}

\begin{proof}[Sketch of proof.]
The operator $P_k$ in \eqref{equ:Cauchy} is specified by
\begin{equation*}
P_{k}(\varphi,v)=\Bigl[ Q_{k}(v \circ \varphi^{-1})\Bigr] \circ
\varphi,
\end{equation*}
where $Q_{k} : C^\infty (\mathbb{S}^{1}) \to C^\infty
(\mathbb{S}^{1})$ is defined by $Q_k(w)=B_k(w,w)+ww_x$. Since
\begin{equation*}
C^\infty (\mathbb{S}^{1})= \bigcap_{r \ge n} H^{r}(\mathbb{S}^{1})
\end{equation*}
for all $n \ge 0$, we may consider the problem \eqref{equ:Cauchy}
on each Hilbert space $H^{n}(\mathbb{S}^{1})$. If $k \ge 1$ and $n
\ge 3$, then $P_k$ is a smooth map from $U^n \times
H^{n}(\mathbb{S}^{1})$ to $H^{n}(\mathbb{S}^{1})$, where $U^{n}
\subset H^{n}(\mathbb{S}^{1})$ is the open subset of all functions
having a strictly positive derivative. The classical
Cauchy-Lipschitz theorem in Hilbert spaces \cite{Lang99} yields
the existence of a unique solution $\varphi_{n}(t) \in U^{n}$ of
\eqref{equ:Cauchy} for all $t \in [0,T_n)$ for some maximal
$T_{n}>0$. Relation \eqref{equ:Momentum} can then be used to prove
that $T_{n}=T_{n+1}$ for all $n \ge 3$.
\end{proof}

\begin{remark}
For $k=0$, in problem \eqref{equ:Cauchy}, we obtain
\begin{equation*}
    P_{0}(\varphi , v) = -2\,\frac{v \cdot v_{x}}{\varphi_{x}}
\end{equation*}
which is not an operator from $U^{n} \times H^{n}(\mathbb{S}^{1})$
into $H^{n}(\mathbb{S}^{1})$ and the proof of
Theorem~\ref{thm:Theorem2} is no longer valid. However, in that
case, the method of characteristics can be used to show that even
for $k=0$ the geodesics exists and are smooth (see \cite{CK02}).
$\lozenge$
\end{remark}

\subsection{The exponential map}

The previous results enable us to define the Riemannian
exponential map $\mathfrak{exp}$ for the $H^k$ right-invariant
metric ($k \ge 0$). In fact, there exists $\delta>0$ and $T>0$ so
that for all $u_0 \in Diff(\mathbb{S}^{1})$ with $\Vert u_0
\Vert_{2k+1}<\delta$ the geodesic $\varphi(t;u_0)$ is defined on
$[0,T]$ and we can define $\mathfrak{exp}(u_0)=\varphi(1;u_0)$ on
the open set
\begin{equation*}
\mathcal{U} = \set{u_0 \in  Diff(\mathbb{S}^{1}):\; \Vert u_0
\Vert_{2k+1} < \frac{2\,\delta}{T} }
\end{equation*}
of $Diff(\mathbb{S}^{1})$. The map $u_0 \mapsto
\mathfrak{exp}(u_0)$ is smooth and its Fr\'{e}chet derivative at zero,
$D\mathfrak{exp}_0$, is the identity operator. On a Fr\'{e}chet
manifold, these facts alone do not necessarily ensure that
$\mathfrak{exp}$ is a smooth local diffeomorphism
\cite{Hamilton82}. However, in~\cite{CK03}, we proved

\begin{theorem}\label{thm:Theorem3}
The Riemannian exponential map for the $H^k$ right-invariant
metric on $Diff(\mathbb{S}^{1})$, $k \ge 1$, is a smooth local
diffeomorphism from a neighborhood of zero on
$Vect(\mathbb{S}^{1})$ to a neighborhood of $Id$ on
$Diff(\mathbb{S}^{1})$.
\end{theorem}

\begin{proof}[Sketch of proof.]
Working in $H^{k+3}(\mathbb{S}^{1})$, we deduce from the inverse
function theorem in Hilbert spaces that $\mathfrak{exp}$ is a
smooth diffeomorphism from an open neighborhood
$\mathcal{O}_{k+3}$ of $0 \in H^{k+3}(\mathbb{S}^{1})$ to an open
neighborhood $\Theta_{k+3}$ of $Id \in U^{k+3}$.

We may choose $\mathcal{O}_{k+3}$ such that
$D\mathfrak{exp}_{u_0}$ is a bijection of
$H^{k+3}(\mathbb{S}^{1})$ for every $u_0 \in \mathcal{O}_{k+3}$.
Given $n \ge k+3$, using \eqref{equ:Momentum} and the geodesic
equation, we conclude that there is no $u_0 \in
H^{n}(\mathbb{S}^{1}) \setminus H^{n+1}(\mathbb{S}^{1}), $ with
$\mathfrak{exp}(u_0) \in U^{n+1}$. We have proved that for every
$n \ge k+3$,
\begin{equation*}
\mathfrak{exp}: \mathcal{O}=\mathcal{O}_{k+3}\,\cap \,
C^\infty(\mathbb{S}^{1}) \rightarrow \Theta=\Theta_{k+3} \,\cap \,
C^\infty(\mathbb{S}^{1})
\end{equation*}
is a bijection. Using similar arguments, \eqref{equ:Momentum} and
the geodesic equation can be used to prove that there is no $u_0
\in H^{n}(\mathbb{S}^{1})\setminus H^{n+1}(\mathbb{S}^{1})$, with
$D\mathfrak{exp}_{u_0}(v) \in H^{n+1}(\mathbb{S}^{1})$ for some
$u_0 \in \mathcal{O}$. Hence, for every $u_0 \in \mathcal{O}$ and
$n \ge k+3$, the bounded linear operator $D\mathfrak{exp}_{u_0}$
is a bijection from $H^{n}(\mathbb{S}^{1})$ to
$H^{n}(\mathbb{S}^{1})$.
\end{proof}

\begin{remark}
For $k=0$ we have that $\mathfrak{exp}$ is not a $C^1$ local
diffeomorphism from a neighborhood of $0 \in Vect(\mathbb{S}^{1})$
to a neighborhood of $Id \in Diff(\mathbb{S}^{1})$, as proved in
\cite{CK02}. The crucial difference with the case ($k \ge 1$) lies
in the fact that the inverse of the operator $A_k$, defined by
\eqref{equ:Ak}, is not regularizing. This feature makes the
previous approach inapplicable but the existence of geodesics can
nevertheless be proved by the method of characteristics.
$\lozenge$
\end{remark}

%--------------------------------------------------------------------
%\nocite{*}
%\bibliographystyle{plain}
%\bibliography{Oberwolfach}
% ----------------------------------------------------------------

\end{document}